\documentclass[twocolumn,showpacs,preprintnumbers,amsmath,amssymb]{revtex4}
\usepackage{graphicx}% Include figure files
\usepackage{dcolumn}% Align table columns on decimal point
\usepackage{bm}% bold math
\usepackage{color}
\usepackage{ulem}

\begin{document}

%\preprint{APS/123-QED}

\title{
Partial Kondo screening in frustrated Kondo lattice systems
}% Force line breaks with \\

\author{Yukitoshi Motome, Kyoya Nakamikawa, Youhei Yamaji, and Masafumi Udagawa}
\affiliation{
 Department of Applied Physics, University of Tokyo, Hongo, Tokyo 113-8656, Japan}%Lines break 

\date{\today}% It is always \today, today,

\begin{abstract}
We investigate the effect of geometrical frustration on the competition 
between the Kondo coupling and the Ruderman-Kittel-Kasuya-Yosida interaction 
in Kondo lattice systems. 
By variational Monte Carlo simulations, 
we reveal an emergent quantum phase with partial ordering 
in which the frustration is relieved by forming a magnetic order on a sublattice 
and leaving the rest in the Kondo screening with spin-singlet formation. 
The role of quantum fluctuations and spin-charge interplay is elucidated. 
\end{abstract}

\maketitle

Kondo lattice systems provide a fertile ground for studying fascinating phenomena 
in strongly-correlated electron systems~\cite{Hewson}. 
The key concept is competition between 
the Kondo coupling and the Ruderman-Kittel-Kasuya-Yosida (RKKY) interaction. 
The former is a local antiferromagnetic (AFM) interaction 
between conduction electrons and localized moments, 
which promotes spin-singlet formation and results in Fermi liquid states,
as the so-called Kondo effect~\cite{Kondo1964}. 
On the other hand, the latter RKKY interaction is an effective magnetic coupling 
between localized moments mediated by conduction electrons, 
which tends to stabilize a magnetic ordering~\cite{Ruderman1954}. 
The competition leads to a quantum critical point (QCP) 
between a Fermi liquid state and a magnetically ordered state~\cite{Doniach1977}. 
QCP has attracted much attention as a source of fascinating phenomena, 
such as non-Fermi-liquid behavior and superconductivity. 

In the present study, we explore yet another phenomenon emergent from 
the competition between the Kondo coupling and the RKKY interaction. 
Our interest is in the possibility to have an intermediate quantum phase 
induced by geometrical frustration, which preempts QCP, 
with a coexistence of screened local moments due to the Kondo singlet formation and
magnetically ordered moments stabilized by the RKKY interaction.
This is a partially ordered state, 
which we call the partial Kondo screening (PKS) state in this paper. 

The partial ordering is sometimes seen in localized spin systems 
with geometrical frustration~\cite{Mekata1977}. 
Our target PKS is, however, qualitatively different 
from the partial ordering in localized spin systems in the following points: 
The system includes itinerant electrons, and 
the disordered sites are not simply paramagnetic 
but participate in the Kondo singlet formation with vanishing their moments. 
These differences will bring about many distinctive aspects 
not only in the stabilization mechanism of the partial order 
but also in the resulting physical properties. 

Several candidates for PKS are experimentally found in $f$-electron compounds, 
e.g., a distorted kagome material CePdAl~\cite{Donni1996,Oyamada2008} 
and a triangular material UNi$_4$B~\cite{Mentink1994,Oyamada2007}. 
These PKS states were theoretically studied 
by the mean-field approximation of a pseudomoment model, 
which describes the magnetic and singlet states by discrete classical variables~\cite{NunezRegueiro1997,Lacroix1996,Ballou1998}. 
In the previous studies, the effects of quantum fluctuations and 
the interplay between conduction electrons and localized moments 
are not fully taken into account, 
despite the fact that they are obviously crucial in the spin-charge coupled systems. 

In this Letter, we explore PKS as a quantum phase for 
the Kondo lattice model and the Kondo necklace model on frustrated lattices. 
The Kondo lattice model (KLM) is one of the fundamental models for rare-earth compounds, 
whose Hamiltonian reads 
\begin{equation}
{\cal H} = -t \sum_{\langle ij \rangle \sigma} (c_{i \sigma}^\dagger c_{j \sigma} +{\rm h.c.}) 
+ J \sum_i \bm{\tau}_i \cdot \bm{S}_i 
+ I_z \sum_{\langle ij \rangle} S_i^z S_j^z,
\label{eq:H_KLM}
\end{equation}
where the first term describes the hopping of conduction electrons and 
the second term represents the Kondo coupling 
between the conduction electron spin $\bm{\tau}_i$ and the localized spin $\bm{S}_i$. 
For simplicity, we consider $S=1/2$ spins for the localized spins. 
In Eq.~(\ref{eq:H_KLM}) we extend the model by adding the last term, 
the AFM Ising interaction between localized spins 
($S_i^z$ is the $z$ component of $\bm{S}_i$), 
in order to mimic the magnetic anisotropy often seen in real materials~\cite{anisotropy}. 
The Kondo necklace model (KNM) is a simplified variant of KLM at half filling~\cite{Doniach1977}: 
\begin{equation}
{\cal H} = W \sum_{\langle ij \rangle} \bm{\tau}_i \cdot \bm{\tau}_j 
+ J \sum_i \bm{\tau}_i \cdot \bm{S}_i 
+ I_z \sum_{\langle ij \rangle} S_i^z S_j^z, 
\label{eq:H_KNM}
\end{equation}
where the charge degree of freedom of conduction electrons is suppressed 
with assuming that there is one electron localized at every site. 
Instability toward PKS in KNM was studied previously by the authors 
for limited lattice geometries and system sizes~\cite{Motome2009}. 
We consider the two models 
on one of the simplest frustrated lattices, the two-dimensional triangular lattice, 
and take the sums $\langle ij \rangle$ over the nearest-neighbor sites. 

We study the ground state of the models (\ref{eq:H_KLM}) and (\ref{eq:H_KNM}) 
by the variational Monte Carlo (VMC) method. 
The method has several advantages compared to others; 
e.g., it takes account of quantum fluctuations neglected in the mean-field approximation, and 
it can avoid the minus sign problem even in frustrated systems. 
We here consider the variational wave function in the form 
\begin{equation}
|\psi \rangle = P_{\rm{G}} {\cal L}^{S=0} {\cal L}^{K=0} |\phi_{\rm{pair}} \rangle, 
\label{eq:psi}
\end{equation}
which describes magnetic states and nonmagnetic singlet states on an equal footing 
as a natural extension of the Yosida-type wave function~\cite{Shiba1990}. 
Here $|\phi_{\rm{pair}} \rangle$ is a generalized BCS wave function defined by 
$
|\phi_{\rm{pair}} \rangle = 
( \sum_{\ell,m=1}^{2N} f_{\ell m} a_{\ell \uparrow}^\dagger a_{m \downarrow}^\dagger)^{N_{\rm e}} |0 \rangle,
$
where $f_{\ell m}$ are the variational parameters and $|0 \rangle$ is a vacuum: 
The fermion operators $a_{\ell\sigma}$ represent 
both conduction $c_{i\sigma}$ and localized electrons (or localized spins), 
resulting in pair creations for any combination of them.
We focus on the half-filling case by setting $N_{\rm e}=N$, 
where $N$ is the number of lattice sites.
The electron number is fixed to be one at each localized spin site. 
${\cal L}^{S=0}$ and ${\cal L}^{K=0}$ are the quantum-number projection operators 
for the total spin singlet and the total momentum zero, respectively.
$P_{\rm{G}}$ is the Gutzwiller factor for optimizing the weight 
of configurations with double occupancies in KLM; 
$P_{\rm{G}} = \exp(-\sum_i g_i n_{i \uparrow} n_{i \downarrow})$
where $g_i$ are the variational parameters and $n_{i \sigma} = c_{i \sigma}^\dagger c_{i \sigma}$. 
We follow Ref.~\cite{Tahara2008} to optimize a large number of variational parameters
by using the stochastic reconfiguration~\cite{Sorella1998} and 
to enforce the quantum-number projections. 

In the present study, we explore the solutions with three-sublattice ordering 
by imposing ${\cal L}^{K=0}$ only for the same sublattices and 
by considering only the sublattice dependence of $g_i$ in $P_{\rm{G}}$. 
${\cal L}^{S=0}$ is used only for $I_z=0$. 
Typically, the optimization is achieved by 300-1000 stochastic reconfiguration steps 
with 1600-6000 Monte Carlo samplings. 
We confirm that the optimized wave function gives a precise ground-state energy
compared to the results by the exact diagonalization; 
e.g., the relative error is less than 0.03 for KNM with $N=12$. 
We apply the method to the clusters with system size $N=12, 18$, and $24$
with imposing the boundary conditions compatible with the three-sublattice order. 

\begin{figure}[t]
\begin{center}
\includegraphics[width=0.4\textwidth]{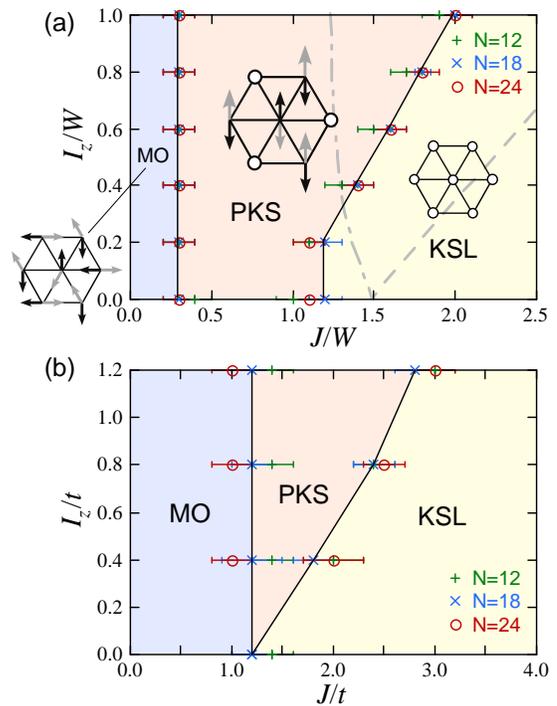}
\end{center}
\caption{\label{fig:phase} 
(color online). 
Phase diagrams for (a) the Kondo necklace model 
and (b) the Kondo lattice model at half filling 
on the triangular lattice determined by VMC. 
The lines are guides for the eye connecting the data for $N=18$. 
There are three phases: MO, PKS, and KSL. 
Schematic pictures for the spin state in each phase are shown in (a), 
where the gray (black) arrows represent $\bm{\tau}_i$ ($\bm{S}_i$) and 
the circles denote the Kondo singlets.
The dot-dashed and dashed lines in (a) show the phase boundaries 
for MO-PKS and PKS-KSL, respectively, determined by the mean-field approximation. 
}
\end{figure}

First we discuss the results for KNM given by Eq.~(\ref{eq:H_KNM}). 
Figure~\ref{fig:phase}(a) summarizes the ground-state phase diagram determined by VMC. 
There are three distinct regions, i.e., 
a magnetically-ordered (MO) state in the small $J/W$ region, 
a Kondo spin liquid (KSL) state in the large $J/W$ region, and a PKS state in between. 
The MO state for small $J/W$ has a three-sublattice ordering 
as schematically shown in the figure. 
This peculiar order is governed by the first and third terms in the Hamiltonian (\ref{eq:H_KNM}) 
as discussed later. 
In the opposite large-$J/W$ region, the second term becomes dominant, and 
the Kondo singlet is formed at every site to realize KSL. 
In the intermediate competing regime, we obtain the PKS phase in which 
one sublattice is  dominated by the local Kondo singlet formation and 
the remaining two retain magnetic ordering, as schematically depicted in the figure. 

\begin{figure}[t]
\begin{center}
\includegraphics[width=0.48\textwidth]{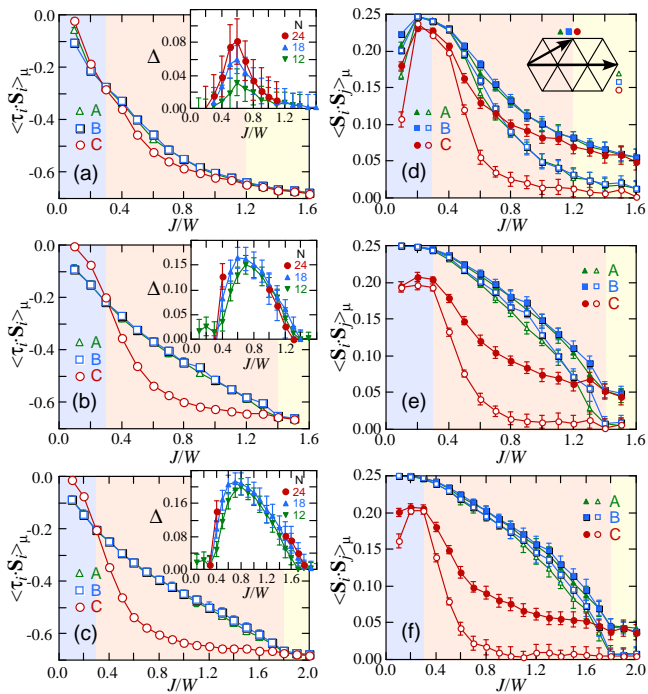}
\end{center}
\caption{\label{fig:KNM_tauS_SS} 
(color online). 
(a)-(c) Onsite correlation for the three-sublattice $\mu=$ A, B, C. 
The insets show their difference $\Delta$ for various system sizes $N$. 
(d)-(f) Intersite correlation between localized spins on the same sublattice. 
Closed and open symbols show the results for the distance $\sqrt{3}$ and $3$, respectively, 
as shown in the inset in (d). 
Data in the main panels are for KNM with $N=18$ and at 
(a),(d) $I_z=0.0$, (b),(d) $I_z/W=0.4$, and (c),(f) $I_z/W=0.8$. 
The lines are guides for the eye.
}
\end{figure}

Then we show how we identify three phases in the following.
Figure~\ref{fig:KNM_tauS_SS} plots $J/W$ dependences of 
the onsite correlation $\langle \bm{\tau}_i \cdot \bm{S}_i \rangle_{\mu}$ and 
the intersite correlation between the localized spins on the same sublattice 
$\langle \bm{S}_i \cdot \bm{S}_j \rangle_{\mu}$ ($\mu$ denotes A, B, or C). 
In the intermediate $J/W$ region, 
the onsite correlation $\langle \bm{\tau}_i \cdot \bm{S}_i \rangle_{\mu}$ on one sublattice 
(C in this case) becomes considerably larger in magnitude than the other two, 
as shown in Figs.~\ref{fig:KNM_tauS_SS}(a)-(c). 
The difference 
$\Delta = | \langle \bm{\tau}_i \cdot \bm{S}_i \rangle_{\rm{C}} 
- (\langle \bm{\tau}_i \cdot \bm{S}_i \rangle_{\rm{A}} 
+ \langle \bm{\tau}_i \cdot \bm{S}_i \rangle_{\rm{B}})/2 |$ is plotted 
for different system sizes $N$ in each inset. 
In the same region, $\langle \bm{S}_i \cdot \bm{S}_j \rangle_{\rm{C}}$ is suppressed 
compared to those for A and B sublattices, and moreover, 
it decays quicker with increasing distance than the other two [Figs.~\ref{fig:KNM_tauS_SS}(d)-(f)]. 
These are clear indications of PKS; 
the local Kondo singlet is enhanced on the C sublattice 
compared to the rest magnetically active sublattices~\cite{note}. 

\begin{figure}[t]
\begin{center}
\includegraphics[width=0.45\textwidth]{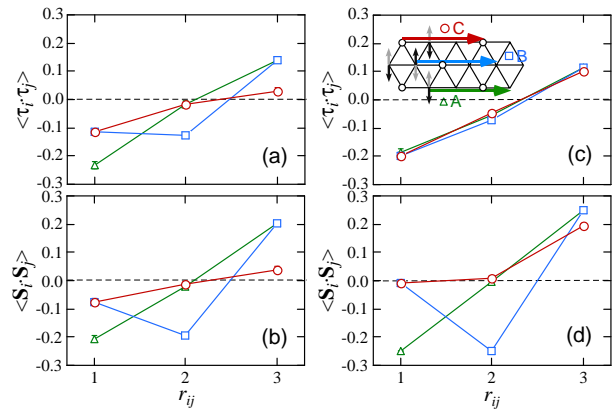}
\end{center}
\caption{\label{fig:KNM_tautau_SS} 
(color online). 
Intersite correlation at $I_z/W=0.4$ for (a),(b) $J/W=0.6$ and (c),(d) $J/W=0.2$
in the KNM with $N=18$.
The correlations are measured along the horizontal arrows in the inset in (c) with fixing the site $i$ 
at the origins of the arrows and shifting $j$. 
}
\end{figure}

Such specific spin configuration in the PKS state is also evidenced 
by the intersite correlations among different sublattices 
plotted in Figs.~\ref{fig:KNM_tautau_SS}(a) and \ref{fig:KNM_tautau_SS}(b). 
For both $\bm{\tau}_i$ and $\bm{S}_i$, 
the intersite correlations measured from the C sublattice are suppressed, 
while those between A and B sublattices are robustly AFM. 
Therefore, the spin configuration of the PKS state is basically composed of 
enhanced Kondo singlets on one sublattice and the AFM network 
on the remaining unfrustrated honeycomb lattice, 
as schematically shown in Fig.~\ref{fig:phase}(a). 

In the smaller $J/W$ region, 
Figs.~\ref{fig:KNM_tautau_SS}(c) and \ref{fig:KNM_tautau_SS}(d) show that 
$\langle \bm{\tau}_i \cdot \bm{\tau}_j \rangle$ is almost independent of the sublattice, 
while $\langle \bm{S}_i \cdot \bm{S}_j \rangle \simeq \pm 1/4$ 
between A and B but $\langle \bm{S}_i \cdot \bm{S}_j \rangle \simeq 0$ 
between C and other sublattices. 
These behaviors indicate the three-sublattice order for the MO phase in Fig.~\ref{fig:phase}(a): 
$\bm{\tau}_i$ forms almost $120^\circ$ order 
to optimize the $W$ term, 
while $\bm{S}_i$ exhibits an almost collinear AFM order 
in A and B sublattices with leaving perpendicular spins on the C sublattice 
to optimize the $I_z$ term. 

On the other hand, for large $J/W$, 
all $\langle \bm{\tau}_i \cdot \bm{S}_i \rangle_{\mu}$ become identical 
within the statistical errorbar and 
take a value close to $-3/4$, as shown in Figs.~\ref{fig:KNM_tauS_SS}(a)-(c). 
At the same time, all $\langle \bm{S}_i \cdot \bm{S}_j \rangle_{\mu}$ merge 
and become small with showing a rapid decay with the distance, 
as plotted in Figs.~\ref{fig:KNM_tauS_SS}(d)-(f). 
These are a sign of the KSL phase where the spins form singlets locally  
as $\langle \bm{\tau}_i \cdot \bm{S}_i \rangle_{\mu} \sim -3/4$. 

The phase boundaries are determined from the behaviors of these spin correlations. 
In particular, the MO-PKS boundary is determined by a sudden change of 
intersite spin correlations in Figs.~\ref{fig:KNM_tauS_SS}(d)-(f) 
as well as by the onset of $\Delta$ in Figs.~\ref{fig:KNM_tauS_SS}(a)-(c). 
Two states have the same symmetry in terms of spins apparently~\cite{note}, 
but we conclude that there is a phase transition between them 
in the light of the mean-field results discussed below. 

Let us make a remark on the limit of $I_z \gg J$ ($\gg W$). 
In this limit, $\vec{S}_i$ are decoupled from $\vec{\tau}_i$ 
and expected to be disordered with macroscopic degeneracy~\cite{Ising}. 
Since $J$ tends to lift the degeneracy through flipping $\vec{S}_i$, 
the system is mapped onto a transverse-field Ising model on the triangular lattice, 
for which a three-sublattice partial order is suggested to appear~\cite{Moessner2001}. 
Such consideration leads us to expect the three-sublattice PKS state 
in the limit of $I_z \gg J$; 
this appears to be consistent with the result in Fig.~\ref{fig:phase}(a). 

The VMC phase diagram is compared with that by the mean-field approximation (MFA) 
in Fig.~\ref{fig:phase}(a). 
Here we perform MFA which accommodates three-sublattice ordering 
by extending the method in Ref.~\cite{Doniach1977}.
MFA also predicts three phases, MO, PKS, and KSL. 
However, there are several differences compared with the VMC results. 
The most distinct one is that 
VMC predicts the PKS state in a finite range of $J/W$ down to $I_z=0$, 
whereas it disappears at $I_z=0$ in MFA. 
Another important difference is that 
VMC phase boundaries shift to smaller $J/W$ compared to the MFA results, and 
the width of MO state becomes relatively narrower.  
These are consequences of intersite quantum fluctuations neglected in MFA.  

\begin{figure}[t]
\begin{center}
\includegraphics[width=0.48\textwidth]{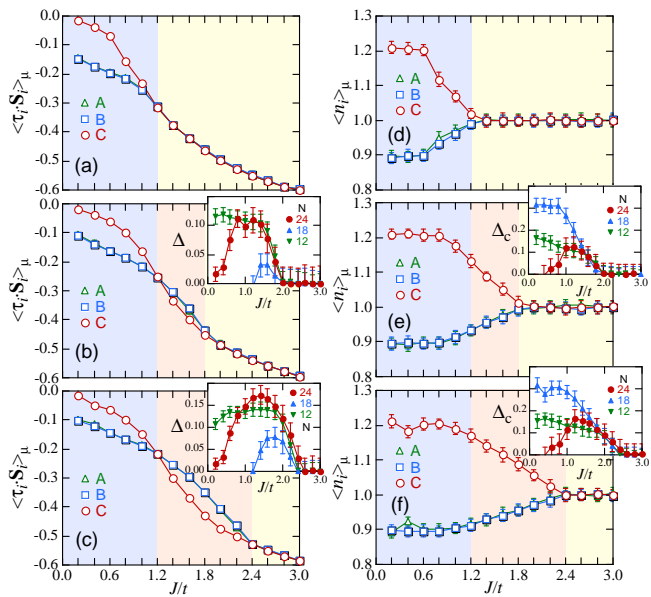}
\end{center}
\caption{\label{fig:KLM_tauS_n} 
(color online). 
(a)-(c) Onsite correlation and (d)-(f) local electron density 
in KLM with $N=18$. 
The insets show their differences with varying $N$.  
The data are for (a),(d) $I_z=0.0$, (b),(d) $I_z/t=0.4$, and (c),(f) $I_z/t=0.8$. 
}
\end{figure}

Now we turn to KLM in Eq.~(\ref{eq:H_KLM}), which includes itinerancy of electrons explicitly. 
Figure~\ref{fig:phase}(b) summarizes the phase diagram at half filling. 
We find that KLM exhibits a similar sequence of three phases including PKS. 
The corresponding onsite correlations are plotted in Figs.~\ref{fig:KLM_tauS_n}(a)-(c). 
As seen in KNM [Figs.~\ref{fig:KNM_tauS_SS}(a)-(c)], 
$\langle \bm{\tau}_i \cdot \bm{S}_i \rangle_\mu$ 
becomes larger on one sublattice than the other two in the PKS region. 
Although $\Delta$ [insets in Fig.~\ref{fig:KLM_tauS_n}(b) and \ref{fig:KLM_tauS_n}(c)] 
suffers from finite-size effects 
(the reason is discussed below), 
we determine phase boundaries from qualitatively similar changes of spin correlations 
to those in Figs.~\ref{fig:KNM_tauS_SS}(d)-(f) and in Fig.~\ref{fig:KNM_tautau_SS} (not shown). 
At $I_z=0$, PKS behavior is not clearly observed in KLM [Fig.~\ref{fig:KLM_tauS_n}(a)]. 
Furthermore, the MO region is wider than in KNM; we return to this point below. 
Consequently, PKS region becomes relatively narrower compared to the KNM case, 
yet remains robust between the MO and KSL phases 
even when conduction electrons are considered explicitly. 

The relatively wider MO phase is presumably attributed to the complicated role of $J$ in KLM:
$J$ enhances the RKKY interaction which tends to stabilize magnetic ordering, 
in addition to the enhancement of the spin-singlet formation. 
In addition, the notable system-size dependence in the small $J/t$ region 
[the insets in Fig.~\ref{fig:KLM_tauS_n}] might be due to 
the long-ranged and oscillating nature of the RKKY interaction~\cite{Ruderman1954}, 
which is difficult to capture within the small finite-size clusters. 
Further study in larger system sizes is desired to clarify the nature of MO in KLM. 

Another interesting observation related with the charge degree of freedom in KLM 
is that PKS accompanies charge disproportionation. 
The local charge $\langle n_i \rangle_\mu = \sum_\sigma \langle n_{i \sigma} \rangle_\mu$ 
disproportionates among the sublattices as shown in Figs.~\ref{fig:KLM_tauS_n}(d)-(f). 
An instability toward charge density wave was recently discussed 
for the unfrustrated KLM model around quarter filling~\cite{Otsuki2009}. 
The relation is not clear between the instability and 
our PKS with charge disproportionation. 
It is interesting to study the possibility of PKS for general filling, 
in particular, at commensurate filling. 

In summary, 
we have investigated the effect of geometrical frustration on 
the competition between the RKKY interaction and the Kondo coupling 
by the variational Monte Carlo simulation for 
the Kondo lattice model and the Kondo necklace model. 
The comparative study between two models reveals the following features.
(i) Both models exhibit PKS phase in between MO and KSL phases. 
(ii) The PKS state is further stabilized by quantum fluctuations and the spin anisotropy. 
(iii) Charge degree of freedom manifests in the stability of the MO phase and 
in the charge disproportionation associated with PKS. 
All the results illuminate the appearance of the PKS phase in the Kondo lattice systems 
even when taking into account quantum fluctuations and spin-charge interplay, 
both of which have not been studied in the previous studies. 
We believe that the results pave the way for further understanding of PKS 
observed in complicated materials and 
of expected spin-charge entangled phenomena inherent to PKS. 
Our results will also cast a new light on the recent efforts 
to explore a new paradigm of QCP physics in the Kondo problem~\cite{Si2006,Custers}. 

The authors thank D. Tahara for the use of his VMC code and useful comments 
and T. Misawa for enlightening discussions. 
This work was supported by KAKENHI (No. 19052008 and No. 21340090), 
Global COE Program ``the Physical Sciences Frontier", 
and by the Next Generation Super Computing Project, Nanoscience Program, MEXT, Japan.


\begin{thebibliography}{9}
\bibitem{anisotropy} 
It might be more realistic to incorporate the single-ion anisotropy for $S \ge 1$. 
To make calculations feasible, we implement the exchange anisotropy for $S=1/2$.
\bibitem{note}
PKS state is better identified by the structure factor for $\langle \bm{\tau}_i \cdot \bm{S}_i \rangle$. 
This requires the calculation of the four-body correlation function such as
$(\bm{\tau}_i \cdot \bm{S}_i) (\bm{\tau}_j \cdot \bm{S}_j)$, 
which is too cpu-time consuming in the present VMC.
\end{thebibliography}
\end{document}